\documentclass[11pt]{article}
\pdfoutput=1
\usepackage[margin=1in]{geometry}
\usepackage[utf8]{inputenc}
\usepackage[T1]{fontenc}
\usepackage{amsmath,amssymb,amsfonts}
\usepackage{bm}
\usepackage{slashed}
\usepackage{physics}
\usepackage{graphicx}
\usepackage{booktabs}
\usepackage{array}
\usepackage{tabularx}
\usepackage{multirow}
\usepackage{makecell}
\usepackage{float}
\usepackage{authblk}
\usepackage[font=small,labelfont=bf]{caption}
\usepackage{subcaption}
\usepackage[dvipsnames]{xcolor}
\usepackage{enumitem}
\usepackage{setspace}
\usepackage{microtype}

\usepackage{fancyhdr}
\fancypagestyle{plain}{
    \fancyhf{}
    \rhead{\small PITT-PACC-2614\\
    FERMILAB-PUB-26-0660-T}
    
}

\usepackage[square,numbers,sort&compress]{natbib}

\usepackage[
  colorlinks=true,
  linktocpage=true,
  linkcolor=DarkOrange,
  citecolor=LighterBlue,
  urlcolor=LighterBlue,
  pagebackref=true
]{hyperref}

\usepackage{hypernat}
\usepackage[nameinlink,noabbrev]{cleveref}

\usepackage{soul}

\definecolor{DarkOrange}{rgb}{0.85,0.30,0.02}
\definecolor{DarkBlue}{rgb}{0.06,0.19,0.32}
\definecolor{PaleBlue}{rgb}{0.91,0.95,0.98}
\definecolor{LighterBlue}{rgb}{0.3,0.3,1}
\definecolor{PaleOrange}{rgb}{1.00,0.96,0.88}
\definecolor{GoodGreen}{rgb}{0.05,0.45,0.30}
\definecolor{GoodYellow}{rgb}{0.95,0.75,0.05}

\crefname{equation}{Eq.}{Eqs.}
\crefname{figure}{Fig.}{Figs.}
\crefname{table}{Table}{Tables}
\crefname{section}{Sec.}{Secs.}

\definecolor{wildstrawberry}{rgb}{1.0, 0.26, 0.64}

\newcommand{\Hparity}{\mathcal H}
\newcommand{\Hp}{\Hparity}

\graphicspath{{figs/}}

\title{For Whom the Xenon Recoils: Magnetic Inelastic Dark Baryons}

\author{Pouya Asadi$^1$, 
Austin Batz$^{2}$, 
Patrick J. Fox$^{3}$, 
Samuel D. Homiller$^{4}$ and 
Graham D. Kribs$^{5}$\\
{\small \color{purple} 
\texttt{pasadi@ucsc.edu,
abatz@caltech.edu, 
pjfox@fnal.gov, 
shomiller@pitt.edu,
kribs@uoregon.edu}
}\\
{\small \textit{${}^1$Department of Physics and Santa Cruz Institute for Particle Physics,} \\ 
\textit{University of California Santa Cruz, Santa Cruz, CA 95064, USA}\\\vskip 0.3cm
\textit{${}^{2}$Walter Burke Institute for Theoretical Physics,} \\ \textit{California Institute of Technology, Pasadena, CA 91125, USA} \\ \vskip 0.3cm
\textit{${}^3$Theory Division, Fermilab, Batavia, IL 60510, USA} \\\vskip 0.3cm
\textit{${}^4$Pitt-PACC, Department of Physics and Astronomy}
\\
\textit{University of Pittsburgh, Pittsburgh, PA, 15260, USA
} \\ \vskip 0.3cm
\textit{${}^{5}$Institute for Fundamental Science and Department of Physics,}\\
\textit{University of Oregon, Eugene, OR 97403, USA} 
}
}

\date{\today}

\begin{document}
    
\maketitle

\vspace{-1.0em}

\begin{abstract}
We develop a theory of a dark baryon dark matter candidate that interacts with nuclei dominantly through inelastic scattering mediated by a transition magnetic dipole operator.  
Models are presented that can elegantly explain both the scattering rate and the size of the inelastic splitting to be consistent with the one event observed at LZ\@. 
Elastic scattering is suppressed due to accidental symmetries within the strongly-coupled sector.
The nuclear response is dominated by a spin-dependent structure factor, thus we find a large range of masses, $1 \; {\rm TeV} \lesssim m_\chi \lesssim 50 \; {\rm TeV}$, and mass splitting of $100 \; {\rm keV} \lesssim \delta \lesssim 500 \; {\rm keV}$ can fit the data for a variety of possible dark matter velocity distributions. 
The models predict that some nuclear scattering events will be accompanied by a simultaneous photon signal with energy $\delta$, which not only distinguishes this model from other inelastic dark matter explanations but also enables directional detection in conventional direct detection experiments. 
We briefly comment on the dark matter abundance, indirect detection signals (including the models' significantly weaker constraints from dark matter annihilation in the Sun), and the associated collider signals that arise from the meson sector of the theory. 
\end{abstract}

\newpage
\tableofcontents

%%%%%%%%%%%%%%%%%%%%%%%%%%%%%%%%%%%%%%%%%%%%%%%%%%%%%%%%%%%%%%%%%%%%%
\section{Introduction}

Direct detection of dark matter (DM) that scatters off nuclei within
exquisitely sensitive underground experiments is among the most sought-after experimental signals of this century.  The recent observation by LZ of one event at large recoil energy that is  consistent with nuclear scattering \cite{LZ:2026axp}, if confirmed with additional data, is a thrilling prospect for uncovering the identity of dark matter. 
The presence of only a single event at high recoil energy may indicate that the scattering of dark matter is occurring  predominantly through an inelastic process, where 
(as the focus of this paper)
dark matter upscatters to an excited state within the dark sector.  

Early searches for inelastic dark matter were limited to relatively low nuclear recoil energies, less than $\sim50$ to $\sim150$~keV, by XENON10 \cite{XENON10:2009sho}, ZEPLIN-III \cite{ZEPLIN-III:2010cnv}, CDMS-II \cite{CDMS-II:2010wvq}, XENON100 \cite{XENON100:2011hxw}.
The importance of a general direct detection search for scattering events at high recoil energies was 
most clearly articulated in
Ref.~\cite{Bramante:2016rdh}
by two of us with Bramante and Martin.
There, it was shown that the focus of xenon experiments 
on relatively low nuclear recoil energies $E_R \lesssim 30$-$50$~keV, limited their  ability to probe large inelasticities 
(while other experiments, such as PICO, did not have such a limitation). 
Since this work, several experimental searches have extended their search region to much larger recoil energies,
including 
XENON100 \cite{XENON:2017pvu,XENON:2017fdd}, 
PandaX \cite{PandaX-II:2017zex,PandaX:2022djq}, 
LUX \cite{LUX:2021ksq},
XENON1T \cite{XENON:2022avm},
PICO \cite{PICO:2023uff},
LZ \cite{LZ:2023lvz,LZ:2026axp},
CDEX-1B \cite{CDEX:2025hez},
XENONnT \cite{XENON:2026tbm},
and future proposals including RES-NOVA \cite{Alloni:2026xdf}.

Theories and phenomenology of inelastic dark matter have a long history.  Early work identified inelastic dark matter candidates arising in weak scale supersymmetric theories \cite{Han:1997wn,Hall:1997ah,Arina:2007tm},
where these ideas were further developed and recognized 
to provide a possible explanation of the DAMA annual modulation signal \cite{TuckerSmith:2001hy,TuckerSmith:2004jv,Chang:2008gd,Cui:2009xq,March-Russell:2008rkh}. 
Composite inelastic theories were also considered in 
\cite{Alves:2009nf,Alves:2010pt} to explain the DAMA annual modulation, where hyperfine splitting within a meson was utilized to obtain the small 
mass splitting between the dark matter and the excited state.
Indirect detection signals
of inelastic dark matter were also found to lead to promising signals \cite{Finkbeiner:2007kk} and additional constraints  \cite{McCullough:2010ai}.  Indeed, indirect detection signals from the Sun have very recently been shown to be quite constraining for certain models \cite{Pospelov:2026ewn}.

Much of this early work capitalized on the idea that a Dirac fermion split by a Majorana mass permits only off-diagonal couplings of a vector mediator to the dark matter state and the excited state.  
Subsequent theories of inelastic dark matter have spanned a wide gamut; 
magnetic inelastic (Majorana) dark matter was the precursor to the ideas presented in this paper
\cite{Masso:2009mu,Chang:2010en,Kumar:2011iy,Patra:2011aa,Weiner:2012cb,Weiner:2012gm,Barello:2014uda}, where the dark matter transitions to an excited state through a magnetic transition moment operator at dimension-5.
Several novel detection strategies arise in this model,
such as looking for \emph{both} the upscatter and the excited state decay to a photon within the detector \cite{Chang:2010en,Lin:2010sb}, which can also be a signal of the model considered in this paper.  Alternatively, 
the excited state decay to photon by itself is also a promising signal
\cite{Feldstein:2010su,Pospelov:2013nea,Eby:2019mgs,Baryakhtar:2020rwy,Bell:2022dbf,Eby:2023wem,Graham:2024syw}. 
A sampling of subsequent work on inelastic dark matter scattering off nuclei includes: heating up neutron stars \cite{Bell:2018pkk,Alvarez:2023fjj};
detection prospects at IceCube \cite{Catena:2018vzc};
detection of large inelastic splittings from
collisional de-excitation
\cite{Pospelov:2019vuf,Lehnert:2019tuw,Broerman:2020hfj,Alves:2023kek,Belli:2025rlx};
new bounds from heavy element nuclear recoil or nuclear excitation
with very low background runs of detectors \cite{Song:2021yar};
production of gamma-rays for indirect detection~\cite{Berlin:2023qco}; formation of compact objects from inelastic dark matter cooling~\cite{Bramante:2023ddr}; direct searches at the Large Hadron Collider (LHC) or beam-dump experiments~\cite{Dienes:2023uve,Asai:2023dzs,Jodlowski:2023ohn,Lu:2023cet}; and capture in the Sun~\cite{Chauhan:2023zuf}.

Strongly-coupled confining theories, that contain dark matter candidates that may scatter inelastically with nuclei, have also been explored extensively (for a recent review, see \cite{Asadi:2026mip} 
by three of us). 
It was discovered in Ref.~\cite{Asadi:2024bbq},
by two of us with Mantel,
that there is a wide class of dark baryon dark matter candidates with highly suppressed elastic scattering cross sections, despite being composed of electrically charged constituents.
The key observation is that this class of theories contains a symmetry, called $\mathcal{H}$-parity, that forbids
one-photon moments, i.e.~magnetic dipole moment, electric dipole moment,
charge radius, and an
anapole moment \cite{Asadi:2024bbq}.
Specific models can contain both $\mathcal{H}$-odd and $\mathcal{H}$-even neutral dark baryons, thereby permitting a magnetic dipole \emph{transition} between these states while all of the single-photon moments vanish for each neutral baryon.

In this paper, we exploit this feature to develop a strongly-coupled theory of magnetic inelastic dark matter (MIDM). 
Strong coupling implies the magnetic dipole transition operator is \emph{not} loop-suppressed, and thus we avoid predicting too-light electrically charged states that have already been ruled out by LEPII and LHC searches.  The mass difference between the neutral composite states can be estimated in the heavy quark limit, as shown in Ref.~\cite{Asadi:2024tpu} by three of us.  In this paper, we show that in certain theories, the mass splittings that are set by electroweak corrections to the inter-quark potential can be $\mathcal{O}(100s)\,$keV\@. 
A magnetic dipole transition operator between dark baryons at these mass splittings is dominated by the spin-dependent nuclear structure factor, which is comparatively large and smooth at large recoil energies $E_R \gtrsim 100$~keV\@.  
We calculate the direct detection scattering rates and nuclear recoil energy spectrum, using both the standard halo model dark matter velocity distribution as well as a model that includes the effects of the Large Magellanic Cloud.  Our central results are shown in \cref{fig:param_scan}.
Taking the recent observation by LZ to motivate the parameter space, we find the scattering rate to be consistent with about one event implies a mass scale of the lightest neutral dark baryon 
in the range $1 \; {\rm TeV} \lesssim m_{\rm DM} \lesssim 50$~TeV with mass splittings $100 \; {\rm keV} \lesssim \delta \lesssim 500 \; {\rm keV}$.  In the heavy quark limit, our calculations of the neutral baryon mass splitting suggest the lower end of this mass range is preferred.  
We conclude with a discussion of  several other predictions and possible constraints on the models, including additional direct detection signals, indirect detection, and collider signals.

Numerous other papers have considered the recent observation by LZ of a single event consistent with dark matter scattering \cite{LZ:2026axp} in a variety of contexts 
\cite{Fan:2026kxx, Wu:2026nhi, Su:2026rwz, Lou:2026idn, Freese:2026sga, 
Yin:2026jnn, Nomura:2026qyq, DiMauro:2026ldr, Visinelli:2026kgt, Yamashita:2026ump,
Chattopadhyay:2026ryw, Smirnov:2026aqk, Du:2026guj, Rodd:2026tyn, McCabe:2026crm, Jeesun:2026vzo, Unwin:2026rdp, 
Dent:2026bji, deLima:2026shq, Gu:2026vto}.

%%%%%%%%%%%%%%%%%%%%%%%%%%%%%%%%%%%%%%%%%%%%%%%%%%%%%%%%%%%%%%%%%%%%%
\section{Magnetic Inelastic Dark Matter From New Confining Sectors}
\label{sec:noblemodels}

The large recoil energy of the LZ event naturally hints at an inelastic dark matter scattering. 
Among the many possible interactions between dark matter, the excited state and nuclei, we are particularly interested in the magnetic dipole transition interaction,\footnote{Depending on CP violation in the dark sector, other electromagnetic moments—particularly an electric dipole moment—may dominate inelastic scattering. Here we focus on the magnetic dipole interaction and leave a comprehensive treatment of other moments to future work.}
\begin{equation}
\label{eq:magnetic_transition_operator}
\mathcal{L} \supset \frac{g_M}{2} \frac{e}{2 m_{\chi}} \bar{\chi}' \sigma_{\mu\nu} \chi F^{\mu\nu}
\end{equation}
where $\chi$ and $\chi'$ are the dark matter and excited state, both Dirac fermions, assumed to have masses $m_{\chi}$ and $m_{\chi} + \delta$, respectively.
The dimensionless coefficient $g_M$ is the ``Bohr magneton'' of this magnetic dipole transition, in units of $e / 2 m_{\chi}$.

For a weakly coupled UV completion, this operator is naturally generated at one loop with $g_M \sim m_{\chi} / (16\pi^2 \Lambda)$, with $\Lambda$ the scale of the UV completion. 
As we will see explicitly in the next section, the cross sections relevant for the LZ event would require
$\Lambda \sim \mathcal{O}(100 \; {\rm GeV})$, implying 
new charged states close to the weak scale, which are severely constrained by searches at colliders. 
On the other hand, if the dark matter and excited state are composites of a new strongly-coupled sector, we expect $g_M \sim 1$, and the new states are at the TeV scale. 

For a generic Dirac fermion, one would naively expect that
the transition magnetic moment in \cref{eq:magnetic_transition_operator} will be accompanied by its elastic counterpart $\sim \bar{\chi} \sigma_{\mu\nu} \chi F^{\mu\nu}$ at the same scale. 
This would dominate the direct detection signal, and give events at much lower recoil energies.
A consistent explanation of the LZ results thus requires some additional structure or mechanism by which these elastic operators are forbidden. 
This is naturally implemented by a Dirac fermion split by a Majorana mass~\cite{Masso:2009mu, Chang:2010en}. Similarly, a Majorana fermion does not have a charge radius, but can have an anapole moment~\cite{Pospelov:2000bq, Fitzpatrick:2010br, Ho:2012bg}. 

In Ref.~\cite{Asadi:2024bbq} a large class of strongly-interacting dark sector models were shown to have vanishing one-photon electromagnetic moments while permitting electromagnetic transition moments between different states.  This arises because of an
accidental ``$\Hp$-parity'' acting on the dark baryons of the theory. 
Also, the lightest spin-1/2 dark baryons are Dirac fermions, that carry a conserved dark baryon number, and so the magnetic transition is qualitatively different from earlier magnetic inelastic models that rely on dark matter being a Majorana fermion.
In the next subsection, we will review the definition and key properties of $\Hp$-parity from Refs.~\cite{Asadi:2024bbq, Asadi:2024tpu}, and then discuss the mass splitting and magnetic transition moment predicted from a particular example of such a strongly-interacting theory.
This example will serve as our benchmark model for realizing the recoil event observed at LZ in the rest of the paper. 

\subsection{Electromagnetic Moments and \texorpdfstring{$\Hp$}{H}-Parity}
\label{sec:Hparity}

We extend the Standard Model with a new confining $\textrm{SU}(N_c)$ sector, where $N_f$ flavors of dark quarks $q$ with bare mass $m_q$ transform in the fundamental representation of $\textrm{SU}(N_c)$.
Our main focus will be on a theory in which the $N_f$ quarks are in an  $N_f$ irreducible representation of Standard Model $\textrm{SU}(2)_L$, specifically $N_c = 3$ and $N_f = 3$, while the discussion below is general. The $\Hp$-parity transformation acts on the dark quarks as a global $\textrm{SU}(2)_L$ transformation of the form
\begin{equation}
    q \stackrel{\Hp\ }{\to} e^{i \pi J_2} \, q\,,
\end{equation}
where $J_2$ is the second generator of $\textrm{SU}(2)_L$ in the $N_f$-dimensional representation. A state in the $(2J+1)$-dimensional representation of $\textrm{SU}(2)_L$ that is composed of dark quarks thus transforms by application of the same operator in the appropriate representation:
\begin{equation} \label{eq:Hmatrixelement}
    \langle m_1|e^{i \pi J_2}|m_2\rangle = (-1)^{J-m_1} \delta_{m_1,-m_2}\,,
\end{equation}
where $m_1,m_2=-J,\dots,J$. The matrix-valued $\textrm{SU}(2)_L$ gauge bosons $W_\mu$ likewise transform as
\begin{equation}
    W_\mu = \frac{\tau^i}{2}\, W_\mu^i \stackrel{\Hp\ }{\to} \left(\exp(i \pi \tau^2/2) \, \frac{\tau^i}{2} \, \exp(-i \pi \tau^2/2)\right)\, W_\mu^i =  \frac{\tau^i}{2} \times
    \begin{cases}
    -W_\mu^i, & i=1,3 \\
    \hphantom{-}W_\mu^i, & i = 2
    \end{cases}\,,
\end{equation}
where $\tau^i$ are the Pauli matrices. This transformation acts identically to charge conjugation on $W_\mu$. Indeed, $U\to e^{i\pi J_2}\,U\,e^{-i\pi J_2} = U^*$ for any $U\in\textrm{SU}(2)$, so $\Hp$ implements complex conjugation on elements of the gauge group. The $\Hp$-parity transformation is also defined to act on the hypercharge gauge boson $B_\mu$ as charge conjugation: $B_\mu\stackrel{\Hp\ }{\to} - B_\mu$. Therefore, in the electroweak-broken phase, the $W^\pm$, $Z$, and electromagnetic field strength tensors transform respectively as
\begin{align}
    W_{\mu\nu}^\pm &\stackrel{\Hp\ }{\to} - W_{\mu\nu
    }^\mp\,, \label{eq:HW}\\
    Z_{\mu\nu} &\stackrel{\Hp\ }{\to} - Z_{\mu\nu}\,, \label{eq:HZ}\\ 
    F_{\mu\nu} &\stackrel{\Hp\ }{\to} - F_{\mu\nu}\,. \label{eq:HF}
\end{align}
Standard Model fields besides the electroweak gauge bosons transform trivially under $\Hp$, so the electroweak interactions explicitly break $\Hp$. However, observable implications of this come with significant loop suppression \cite{Asadi:2024bbq}.

We are interested in neutral dark baryons $\chi$, the lightest of which is our DM candidate. One can see from \cref{eq:Hmatrixelement} that $\chi\stackrel{\Hp\ }{\to}(-1)^J \chi$, so these baryons are eigenstates of $\Hp$, and their $\Hp$-parity depends on the dimension of the electroweak multiplet of which they are the neutral component. Neutral baryons in singlets, 5-plets, 9-plets, etc.~are $\Hp$-even while those in 3-plets, 7-plets, etc.~are $\Hp$-odd. Either way, operators of the following schematic form are $\Hp$-odd:
\begin{equation}
    F^{\mu\nu}\, \overline{\chi}\, \mathcal{O}_{\mu\nu}\, \chi\,,
\end{equation}
where $\mathcal{O}_{\mu\nu}$ contains Lorentz structure (possibly including derivatives). Thus, such operators are forbidden, and na\"ively stringent direct detection constraints from elastic single-photon exchange are relaxed. Higher-order electromagnetic moments such as polarizability are not forbidden, but the corresponding signals are highly suppressed. 

More interestingly for this work, operators of the following form are also allowed:
\begin{equation}
    F^{\mu\nu}\, \overline{\chi}'\, \mathcal{O}_{\mu\nu}\, \chi\,,
\end{equation}
where $\chi$ and $\chi'$ are different neutral dark baryons with \textit{opposite} $\Hp$-parity. Thus, electromagnetic transitions between neutral baryon species are allowed. In principle, as long as the mass splitting between $\chi$ and $\chi'$ is sufficiently small, a dark baryon DM candidate may scatter inelastically in direct detection experiments.
In the upcoming section, we study the mass splitting in benchmark models. 

\subsection{Mass Splittings in the Minimal Models}
\label{sec:mass_splitting}

Dark baryon masses are non-perturbative quantities, so their estimation within perturbation theory comes with irreducible uncertainties. However, when the dark sector confinement scale $\Lambda_\chi$ is much smaller than the dark quark mass $m_q$, the quarks are only loosely bound and non-relativistic, so the dark strong coupling $\alpha_\chi$ at the characteristic energy scale is small. In this regime, we can estimate the mass splittings between baryons in different electroweak representations using the quark model~\cite{DeRujula:1975qlm, Manohar:1983md, Georgi:1984zwz}, where one describes the bound state with an inter-quark potential induced by exchange of dark gluons and electroweak bosons. 
We make use of the techniques developed in Ref.~\cite{Asadi:2024tpu} for this purpose. 

The potentials are Coulomb-like (or Yukawa-like in the cases of the massive $W^\pm$ and $Z$), with corrections such as fine- and hyperfine-structure appearing at higher order in the non-relativistic expansion. For the electromagnetic contribution, the potential between constituents $i$ and $j$ at leading order in the non-relativistic expansion (i.e.~the Fermi-Breit potential) has the form 
\begin{equation} \label{eq:VEM}
    V^{\rm EM}_{ij} = \alpha \, Q_iQ_j \left[\frac{1}{r} 
    -\frac{\pi}{m_q^2}\left(1+\frac{8}{3}\mathbf{S}_i\cdot \mathbf{S}_j\right) \delta^{(3)}(\mathbf{r}) 
    + \frac{1}{2m_q^2}\left(\frac{1}{r} \mathbf{\nabla}_i\cdot\mathbf{\nabla}_j + \frac{1}{r^3} \mathbf{r}\cdot(\mathbf{r}\cdot\mathbf{\nabla}_i)\mathbf{\nabla}_j \right)
    \right],
\end{equation}
where $\alpha$ is the electromagnetic coupling, $Q$ is the electric charge, $\mathbf{S}$ is the spin, and $\mathbf{r}$ is the displacement between the constituents. The second term contains the familiar Darwin term from the fine-structure correction and spin-spin coupling from the hyperfine correction. The third term contains orbit-orbit couplings. Not shown are terms that vanish in the absence of orbital angular momentum. See Ref.~\cite{Asadi:2024tpu} for the $W^\pm$ and $Z$ potentials, including finite mass effects derived from Ref.~\cite{DeSanctis:2009zz}. To estimate the ground state of the mass Hamiltonian, we employ a variational method, minimizing the energy with respect to parameters of a trial spatial wavefunction (in our case, a generalized decaying exponential). Strictly speaking, the variational method only provides an upper bound on the ground state energy, but with a sufficiently well-chosen trial wavefunction, that upper bound serves as a sensible estimate of the true value. Ref.~\cite{Asadi:2024tpu} contains the details of this calculation.

\begin{figure}[t]
    \centering
    \includegraphics[width=0.45\linewidth]{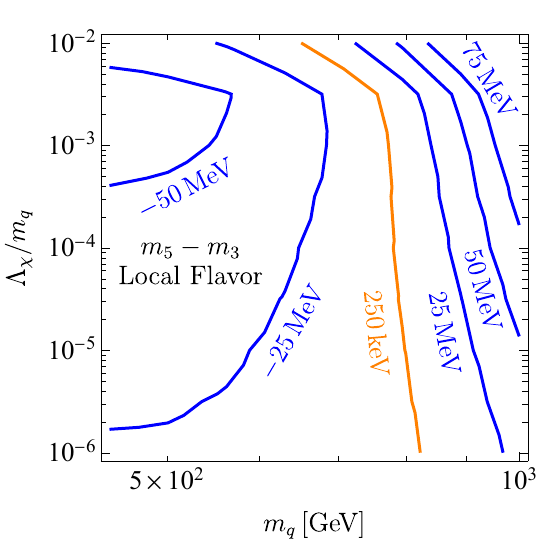}
    \hspace{1.5em}
    \includegraphics[width=0.45\linewidth]{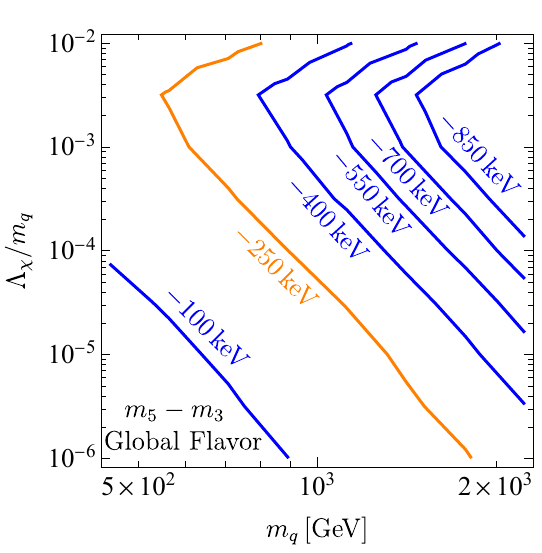}
    \caption{Estimates of the splitting between the neutral SU(2)$_L$ 3-plet and 5-plet dark baryon, with masses $m_3$ and $m_5$, respectively. \textbf{Left:} Our default model with dark quarks in a 3-plet of Standard Model $\textrm{SU}(2)_L$ (i.e.~a local flavor symmetry). 
    An $\mathcal{O}(100)\,$keV mass splitting is achievable (near the \textcolor{DarkOrange}{orange} line). \textbf{Right:} A model with dark quarks charged under $\textrm{U}(1)_Y$ with a global $\textrm{SU}(3)$ flavor symmetry and an analogous $\Hp$-symmetry. Small mass splittings exist over a broader range, but there are additional considerations required to ensure the model is viable. In the upper regions of the plots, where the contours reverse direction, these perturbative estimates become less reliable, as the dark strong coupling is at least $\mathcal{O}(0.1)$.}
    \label{fig:massSplitting}
\end{figure}

We take as an explicit example the case where $N_c = N_f = 3$, i.e.~a dark SU(3) gauge group with three dark quarks transforming as a triplet of SU(2)$_L$. 
This is the simplest model with the dark quarks transforming under SU(2)$_L$, as there are only two neutral baryons in the spectrum \cite{Asadi:2024tpu},\footnote{More precisely, other neutral dark baryons are excited states with higher spin and/or orbital angular momentum.} and they have opposite $\Hp$-parity. They come from a 3-plet and a 5-plet of $\textrm{SU}(2)_L$ and are denoted $B_3^0$ and $B_5^0$, respectively.\footnote{Notably, $\langle B_3^0 | V | B_5^0 \rangle = 0$ for the full potential $V$, even in the electroweak-broken phase, meaning these states do not mix. As discussed in Ref.~\cite{Asadi:2024tpu}, baryons with equal charge in different electroweak representations generically mix, especially when the dark sector mass scales are not very large compared to the electroweak scale. This means classifying states by $\textrm{SU}(2)_L$ representation is not necessarily well-defined. In the case of the $B_3^0$ and $B_5^0$, the mixing is forbidden by the fact that they  have opposite $\Hp$-parity, which is not true of all neutral baryons for different values of $N_c$ and $N_f$. These baryons thus retain their identities as ``3-plet'' and ``5-plet'' insofar as their spin-flavor wavefunctions are unchanged by electroweak symmetry breaking.}  
In this particular case, we find by explicit calculation with the spin-flavor wavefunctions that the mass splitting $m_5-m_3$ between the $B_3^0$ and $B_5^0$ due to the photon and $Z$ are exclusively \textit{hyperfine}, while the splitting due to the $W^\pm$ exists at $0^{\rm th}$ order in the non-relativistic expansion.\footnote{See Ref.~\cite{Asadi:2024tpu} for discussion of the spectrum of dark baryon electroweak representations and spin-flavor wavefunctions for general $N_c$ and $N_f$.}
We find empirically that to a good approximation, the splittings are given by 
\begin{align}
    \Delta_\gamma &\simeq -\frac{\alpha}{m_q^2} \frac{8\pi}{3} \sum_{i<j}\left\langle \delta^{(3)}(\mathbf{r}) \right\rangle, \label{eq:deltagamma}\\
    \Delta_Z &\simeq \frac{2 \alpha_W \cos^2 \theta_W}{3 m_q^2}  \sum_{i<j}\left\langle m_Z^2 \frac{e^{-m_Z r}}{r} -4 \pi \delta^{(3)}(\mathbf{r}) \right\rangle, \label{eq:deltaZ} \\    \Delta_W &\simeq 2\alpha_W
    \sum_{i<j}
    \left\langle
    \frac{e^{-m_W r}}{r}
    + \frac{1}{2m_q^2}
    \frac{e^{-m_W r}}{r^3}
    \mathbf{r}
    \cdot(\mathbf{r}\cdot\mathbf{\nabla}_i)\mathbf{\nabla}_j
    \right\rangle, \label{eq:deltaW}
\end{align}
where $\Delta_\gamma$, $\Delta_Z$, and $\Delta_W$ are, respectively, contributions to $m_5-m_3$ due to the photon, $Z$, and $W^\pm$, $\alpha_W$ is the $\textrm{SU}(2)_L$ fine-structure constant, $\theta_W$ is the Weinberg angle, and $m_{W,Z}$ are the $W^\pm$ and $Z$ masses.  The expectation value is taken with respect to the spatial wavefunctions, which are approximately the same for both baryons. Note that the second (orbit-orbit) term in \cref{eq:deltaW} is negative-definite. In particular, $\Delta_W$ can \textit{change sign} at some value of $m_q$. As  hyperfine corrections, $\Delta_\gamma$ and $\Delta_Z$ are also small, and  $\Delta_Z$ includes a term that is suppressed by $m_Z^2 c_W^2/m_q^2$ that partially cancels the the $Z$ contribution.
As shown in \cref{fig:massSplitting}, the combination of these effects leads to an $\mathcal{O}(100)\,$keV near $m_q=800\,$GeV (corresponding to a DM mass around $2.4\,$TeV).  Typical mass splittings in the $\mathcal{O}(1)$\,TeV DM mass range tend to be $\mathcal{O}(10)\,$MeV or greater, so this calculation, in conjunction with an observation of inelastic DM scattering, predicts the DM mass within a narrow range. For this model, either the $B_3^0$ or the $B_5^0$ could be the DM candidate, as there is parameter space permitting either mass hierarchy.

As an alternative to the model we have been working with thus far, we could consider the case where the dark quark flavor symmetry is purely a global $\textrm{SU}(3)$, and the dark quarks are charged under $\textrm{U}(1)_Y$ rather than $\textrm{SU}(2)_L$ such that they still have $+1$, $0$, and $-1$ electric charge in the electroweak-broken phase:
\begin{equation}
    D_\mu q \supset i g_Y B_\mu 
    \begin{pmatrix}
        \ \, 1 & & \\ & 0 & \\ & & \!-1
    \end{pmatrix}
    \begin{pmatrix}
        q_+ \\ q_0 \\ q_-
    \end{pmatrix}\,,
\end{equation}
where $g_Y$ is the hypercharge coupling, and subscripts denote the electric charge of each flavor. In this case, there is a \textit{global} $\textrm{SU}(2)$ embedded in the flavor symmetry, and there exists an exact analog to $\Hp$-parity wherein the dark quarks transform as
\begin{equation}
    q\stackrel{\Hp\ }{\to} \exp
    \left(\frac{i\pi}{\sqrt{2}}
    \begin{pmatrix}
        0 & -i & 0 \\
        i & 0 & -i \\
        0 & i & 0
    \end{pmatrix}
    \right)
    q
    =
    \begin{pmatrix}
        q_- \\ -q_0 \\ q_+
    \end{pmatrix}
\end{equation}
and the electroweak bosons transform as in \cref{eq:HW,eq:HZ,eq:HF}. Symmetry under this $\mathbb{Z}_2$ forbids diagonal electromagnetic moments while permitting transitions as above.
There is no gauge symmetry forbidding mass splittings among the dark quarks, and indeed these splittings will be radiatively generated, but such symmetry breaking need only be small enough as to not re-introduce problematically large diagonal electromagnetic moments. 

The key feature of this alternative model with global flavor is that single $W^\pm$ exchange in the inter-quark potential is absent, so it does not induce mass splittings between the $B_3^0$ and the $B_5^0$. Instead, those splittings are only induced by hyperfine corrections in \cref{eq:deltagamma,eq:deltaZ}. The implication is that the mass splitting is characteristically small compared to the case where a subgroup of the flavor symmetry is gauged. The fact that the electromagnetic splitting is strictly hyperfine is a special and convenient feature of the $N_c=N_f=3$ model. As shown in the right-hand panel of \cref{fig:massSplitting}, there is a much broader mass range over which the splitting is $\mathcal{O}(100)\,$keV, still in the regime with a TeV-scale DM candidate, compared to the gauged case. There are several potential additional considerations for making this model of global flavor viable, e.g.~introducing higher-dimensional operators to facilitate decay of the charged dark mesons, but we omit further discussion in this work and simply point out that there are modifications to the model we started with that reduce the quark mass tuning required to achieve inelastic DM. 

It is worth emphasizing the uncertainties present in our calculations of the baryon mass splitting. There are non-perturbative corrections, higher-order electroweak corrections, and error introduced by our use of the variational method to approximate the ground state of the mass Hamiltonian. For example, simplifying our trial wavefunction to only a decaying exponential shifts the value of $m_q$ where one achieves an $\mathcal{O}(100)\,$keV splitting by about a factor of two. The magnitude of the splitting is a factor of $\sim\!\mathcal{O}(10^{7})$ smaller than the DM mass scale, so our predictions of the correct DM mass that achieves the desired splitting should not be understood as precise. Rather, we conclude that it is \textit{highly plausible} for the dark baryons to have a sufficiently small mass splitting with an $\mathcal{O}(1)\,$TeV-scale DM candidate. Only a non-perturbative technique such as the lattice could make the prediction precise. Note also that we work in the heavy-quark limit simply because perturbative estimates are feasible, and there is no reason to expect that $\mathcal{O}(100)\,$keV mass splittings cannot be achieved in the light-quark limit. 

Across these two models, mass splittings of \(\delta\sim\mathcal{O}(100)\,\mathrm{keV}\) can be realized for dark-quark masses between \(0.3\) and \(2.0\,\mathrm{TeV}\), corresponding to dark-baryon dark matter masses of \(m_\chi\simeq1\!-\!6\,\mathrm{TeV}\). For suitable values of the transition magnetic dipole moment, dark matter in this mass range can therefore undergo inelastic scattering in the recoil-energy region relevant to the LZ high-recoil anomaly. We next discuss the origin and expected magnitude of this transition magnetic dipole moment before turning to the resulting direct-detection signal.

\subsection{Magnetic Dipole Moments in the Quark Model}

In the non-relativistic limit, the magnetic dipole interaction is governed by the Hamiltonian
\begin{equation}
    H_{\rm mag} = -\boldsymbol{\mu}\cdot\mathbf{B}\,,
\end{equation}
where $\mathbf{B}$ is an external magnetic field, and $\boldsymbol{\mu}$ is the magnetic dipole moment. To estimate the dipole moment $\boldsymbol{\mu}_\chi$ of our dark baryons $\chi$, we use the quark model approximation \cite{Parreno:2016fwu}
\begin{equation} \label{eq:muchi}
    \boldsymbol{\mu}_\chi |\chi\rangle \simeq \sum_{i=1}^{N_c} \boldsymbol{\mu}_i |\chi\rangle = \sum_{i=1}^{N_c} \frac{e}{m_i}\, Q_i\, \mathbf{S}_i |\chi\rangle\,,
\end{equation}
where the state $|\chi\rangle$ contains the constituent spin-flavor wavefunction, $e\simeq0.3$ is the electromagnetic coupling, $\boldsymbol{\mu}_i$ is the dipole moment of the $i^{\rm th}$ constituent with charge $Q_i$ and spin $\mathbf{S}_i$, and $m_i$ is a ``constituent mass'' that is \textit{defined} via $m_i=m_\chi/N_c$ with $m_\chi$ the baryon mass (including in the light-quark limit).

Given the spectrum of dark baryon spin-flavor wavefunctions, we can use \cref{eq:muchi} to compute their magnetic dipole moments, including transition moments. For the $N_c=N_f=3$ model, applying this operator to the spin-flavor wavefunctions yields (with $\boldsymbol{\mu}$ and $\mathbf{B}$ in the $z$-direction)
\begin{equation}
    \Bigg(
    \begin{matrix}
    \langle B_3^0 | \\ \langle B_5^0 |
    \end{matrix}
    \Bigg)\,
    \mu_\chi\,
    \bigg(
    \begin{matrix}
    | B_3^0 \rangle & | B_5^0 \rangle
    \end{matrix}
    \bigg) = \frac{e}{2m_i}
    \begin{pmatrix}
        0 & -\dfrac{2}{\sqrt{3}} \\
        -\dfrac{2}{\sqrt{3}} & 0
     \end{pmatrix}
     =
     \frac{e}{2m_\chi}
    \begin{pmatrix}
        0 & -2\sqrt{3}\ \\
        -2\sqrt{3} & 0
     \end{pmatrix}.
\end{equation}
The vanishing of the diagonal moments can be understood as a consequence of $\Hp$-parity, as explained in \cref{sec:Hparity}. Thus, if we express the dipole moment in terms of the conventional coefficient $g_M$:
\begin{equation}
    \mu_\chi = g_M \frac{e}{2m_\chi}\,,
\end{equation}
we see that the transition moment has
\begin{equation} 
    g_M = -2\sqrt{3} \simeq -3.5\,,
    \label{eq:g_M_numerical}
\end{equation}
roughly double that of the Standard Model neutron magnetic moment. This calculation allows us to estimate the rate of inelastic DM scattering in the following sections.

%%%%%%%%%%%%%%%%%%%%%%%%%%%%%%%%%%%%%%%%%%%%%%%%%%%%%%%%%%%%%%%%%%%%%
\section{Magnetic Inelastic Dark Matter and the LZ Event}

In this section, we turn to the inelastic scattering signal arising from the operator \cref{eq:magnetic_transition_operator}, with an eye towards the composite dark matter models discussed above. 
In the remainder of this section, we  use $g_M = -2 \sqrt{3}$ for the transition moment as in \cref{eq:g_M_numerical} for concreteness. 
First, let us discuss the velocity distribution of dark matter in the halo.
As a default, we assume the Standard Halo Model (SHM) for the this profile in the Milky Way, 
\begin{equation}
f_{\textrm{SHM}}(v)
\equiv
\frac{
e^{-w(v)^2/v_c^2}
}{
(v_c\sqrt{\pi})^3 N_{\mathrm{esc}}
}
\Theta\!\left(v_{\mathrm{esc}}-w(v)\right) \, ,
\label{eq:shm-distribution}
\end{equation}
where the Heaviside function $\Theta$ and normalization factor 
\begin{equation}
\label{eq:shm-normalization}
N_{\mathrm{esc}}
\equiv
\operatorname{erf}\!\left(\frac{v_{\mathrm{esc}}}{v_c}\right)
-
\frac{2}{\sqrt{\pi}}
\frac{v_{\mathrm{esc}}}{v_c}
e^{-v_{\mathrm{esc}}^2/v_c^2} \, ,
\end{equation}
are included to account for the local galactic escape velocity. 
Following Ref.~\cite{DelNobile:2021wmp}, we use $v_{\mathrm{esc}}  \sim 533~\textrm{km/s}$ for the escape velocity and the local circular velocity of the earth is fixed to $v_c = 220~\textrm{km/s}$. $w(v)$ is the DM velocity in the galactic rest frame, 
\begin{equation}
\label{eq:galactic-lab-velocity}
w(v)^2 = v^2+v_{\mathrm{E}}^2
+2v\,v_{\mathrm{E}}\cos\vartheta \, .
\end{equation}
Here, $v_E(t)$ is the Earth velocity with respect to the galactic rest frame, 
\begin{equation}
\label{eq:earth-speed}
v_{\mathrm{E}}(t) =
\sqrt{ v_{\mathrm{S}}^2+v_{\oplus}^2 +2b\,v_{\mathrm{S}}v_{\oplus}
\cos\!\left[ \frac{2\pi}{T_y} (t-t_0)\right]
} \, ,
\end{equation}
and $\vartheta$ denotes the angle between the dark matter velocity $v$ and the Earth galactic velocity $v_E$ in the lab rest frame.
$v_S$ ($v_\oplus$) denote the Solar velocity in the galactic rest frame (Earth velocity with respect to Sun), and $T_y$ is the earth rotation period around the Sun (a sidereal year). 
Rather than integrating the time-dependent velocity distribution over the year, we evaluate all results at the extrema \(b= 1/2\) and with the $\cos t$ factor set to $\pm 1$, thereby bracketing the annual modulation of the signal. We find that the resulting variation is small, especially compared with the uncertainty in the velocity profile itself.

\begin{table}[t]
    \centering
    \renewcommand{\arraystretch}{1.3}
    \begin{tabular}{c|c|c|c|c|c}
        ~ & Weight & $v_c$ & $v_S$  & $v_\oplus$ & $v_\mathrm{esc}$ \\
        \hline
        \hline
       $f_{\textrm{SHM}}$ & 1 & 220 & 232 & 30 & 533 \\
       \hline
       \hline
       $f_{\textrm{SHM}}^{(1)}$ & 0.986 & 218 & 303 & 30 & 473 \\
       \hline
       $f_{\textrm{SHM}}^{(2)}$ & 0.014 & 646 & 12 & 30 & 947
    \end{tabular}
    \caption{Numerical value (in \textbf{km/s}) of different velocities used in the SHM profile in our calculation. The first row corresponds to the standard SHM (see Ref.~\cite{DelNobile:2021wmp} for numerical values), while the other rows are the parameters for two modified SHM-like distributions that we use to fit the effects of the LMC; the net effect of LMC on the velocity profile in given by \cref{eq:f_LMC}.
    }
    \label{tab:SHMvals}
\end{table}

Because of the kinematic threshold, the inelastic scattering rate is very sensitive to the high-velocity tail of the distribution.
This is subject to various significant sources of astrophysical uncertainties. 
In particular, non-equilibrium processes can significantly change the high-velocity tail of the dark matter distribution. Refs.~\cite{Besla:2019xbx, Donaldson:2021byu, Smith-Orlik:2023kyl} argue that interactions between the Large Magellanic Cloud (LMC) and the Milky Way (MW) especially increase the high-speed tail of the local dark matter velocity distribution.

To assess the significance of these effects, we consider in addition to $f_{\textrm{SHM}}$ a modified distribution $f_{\textrm{LMC}}$, which is constructed to include the additional, high-velocity particles.
As a rough approximation, we construct $f_{\textrm{LMC}}$ by summing two 
SHM-like distributions: 
\begin{equation}
f_{\textrm{LMC}}(v) \equiv \kappa f_{\textrm{SHM}}^{(1)}(v) + (1 - \kappa) f_{\textrm{SHM}}^{(2)}(v) \,, 
\label{eq:f_LMC}
\end{equation}
with parameters fitted to mimic the halo integral $\eta_{\textrm{min}}$ given in Ref.~\cite{Smith-Orlik:2023kyl},
\begin{equation}
\eta_{\textrm{min}} = \int_{v_{\textrm{min}}} \dd^3 v \,\frac{1}{v} f(v)\, .
\end{equation}
The numerical values used for different parameters in $f_\mathrm{SHM}^{(1,2)}$ are included in \cref{tab:SHMvals}; it should be emphasized again that $f_\mathrm{SHM}^{(1,2)}$ are not physical distributions, but rather fits to capture the effect of LMC on the dark matter velocity distribution as calculated in Ref.~\cite{Smith-Orlik:2023kyl}.

\begin{figure}[t]
    \centering
    \includegraphics[width=0.7\linewidth]{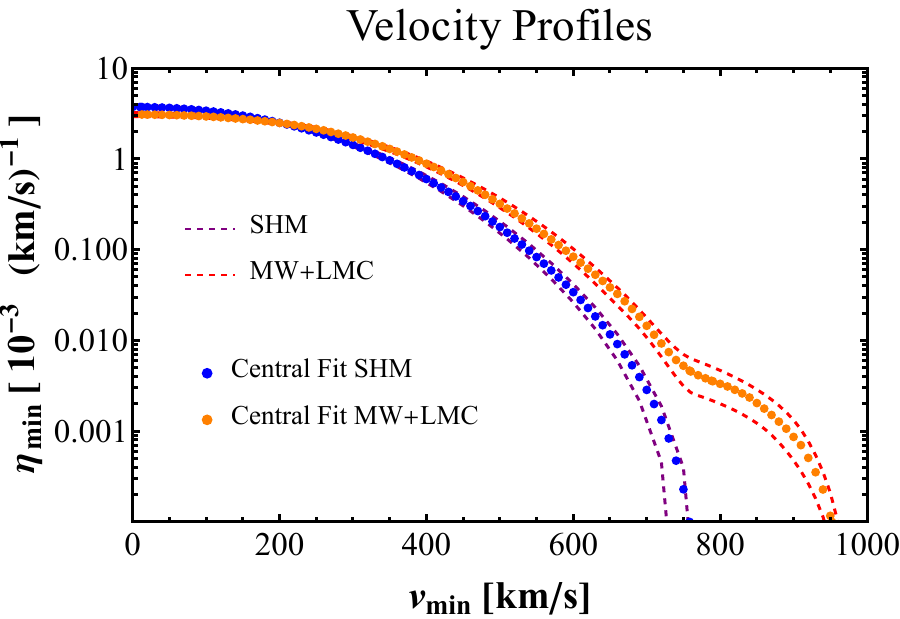}
    \caption{The standard SHM (\textcolor{blue}{blue} dots) and the MW+LMC effect on the dark matter profile (\textcolor{orange}{orange} dots) from Ref.~\cite{Smith-Orlik:2023kyl}. We model the LMC profile as the sum of two unphysical SHM-like distributions, see \cref{eq:f_LMC} and \cref{tab:SHMvals}. The \textcolor{violet}{purple} (\textcolor{red}{red}) dashed lines indicate the two extremes of annual modulation. Comparatively, the annual modulation introduces a far smaller variation in the signal than the choice of the profile.  }
    \label{fig:SHM}
\end{figure}

These velocity profiles are illustrated in Fig.~\ref{fig:SHM}, where we overlay our halo function from $f_{\textrm{SHM}}$ and $f_{\textrm{LMC}}$ with those taken from Ref.~\cite{Smith-Orlik:2023kyl}.
Coincidentally, the perturbation due to the LMC points in the opposite direction of the Sun's motion, and thus the LMC component peaks at a similar time of year as the Standard Halo Model~\cite{Besla:2019xbx}. We therefore neglect any phase difference between these components in our calculation. 
We bracket the annual modulation of the LMC profile similar to the SHM case using $b=1/2$ and setting the $\cos$ factor in \cref{eq:earth-speed} to $\pm 1$, as mentioned above.\footnote{Applying the SHM procedure to the MW+LMC distribution produces only a narrow modulation band at high velocities. To remain conservative, we instead use the uncertainty in the MW+LMC profile—shown by the orange band in Fig.~11 of Ref.~\cite{Smith-Orlik:2023kyl}—as a proxy for annual modulation in the high-velocity tail. This variation is comparable to the modulation estimated in Ref.~\cite{Besla:2019xbx} and likely overestimates its effect.} 
We clearly see that the choice of which profile to use introduces far larger uncertainty than the annual modulation effects.

The rest of the inelastic scattering rate calculation is standard (see e.g.~Ref.~\cite{DelNobile:2021wmp} and references therein). The inelastic scattering off the nucleus follows from Ref.~\cite{Eby:2023wem} (see also Ref.~\cite{Barello:2014uda}). 
The cross section for the inelastic scattering via a magnetic dipole interaction can be obtained with a simple modification of the elastic scattering cross section (see e.g.~Ref.~\cite{DelNobile:2021wmp}, Eq. (H.139)) to account for the effect of a small mass splitting $\delta$. We find,
\begin{equation}
\begin{aligned}
\frac{\mathrm{d}\sigma_T}{\mathrm{d}E_R}
={}&
\frac{e^4}{32\pi} \frac{g_M^2}{m_{\chi}^2} \frac{m_T}{m_{\mathrm{N}}^2 v^2}
\\
&\times
\Bigg[
\frac{2m_{\mathrm N}^2}{m_T}
\left(
\frac{v^2}{E_R}
-\frac{m_{\chi}+2m_T}{2m_{\chi} m_T}
-\frac{\delta(m_{\chi}+m_T)}{m_{\chi} m_T E_R}
-\frac{\delta^2}{2m_T E_R^2}
\right)
F_M^{(p,p)}(q^2)
\\
&\qquad
+4F_{\Delta}^{(p,p)}(q^2)
-2\sum_{N=p,n}
g_N F_{\Sigma'\Delta}^{(N,p)}(q^2)
+\frac{1}{4}\sum_{N,N'=p,n}
g_N g_{N'} F_{\Sigma'}^{(N,N')}(q^2)
\Bigg] \, ,
\end{aligned}
\label{eq:mdm-inelastic-cross-section}
\end{equation}
where the nuclear recoil energy is $E_R = q^2 / (2 m_T)$ with $q$ the momentum transfer and $m_T$ is the mass of the nucleus. 
The average mass of a nucleon is $m_{\textrm{N}} = 939~\textrm{MeV}$, and $g_N$ is the $g$-factor of the nucleon $n$, with 
\begin{equation}
g_p =+5.59 \,, \qquad g_n = -3.83. 
\end{equation}
The ``squared form factors'' $F_{X,Y}^{(N,N')}$, $F_X^{(N,N')} \equiv F_{X,X}^{(N,N')}$ are defined as in Ref.~\cite{Fitzpatrick:2012ix}; we use the results of Ref.~\cite{Anand:2013yka} for all nuclear form factors.
The prefactor includes the factors of the Bohr magneton written in \cref{eq:magnetic_transition_operator}.

The event rate was computed using both our own implementation of the integration, as well as with the WimPyDD package~\cite{Jeong:2021bpl}, and the results corroborate one another. 
We also checked that the individual operator coefficients are in agreement with the calculation in Ref.~\cite{Eby:2023wem}. 

\begin{figure}[t]
    \centering
    \resizebox{\columnwidth}{!}{\includegraphics[width=0.5\linewidth]{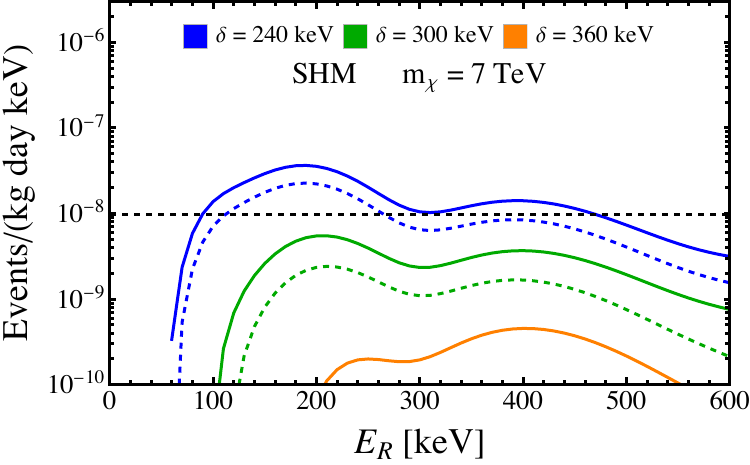}
    \quad
    \includegraphics[width=0.5\linewidth]{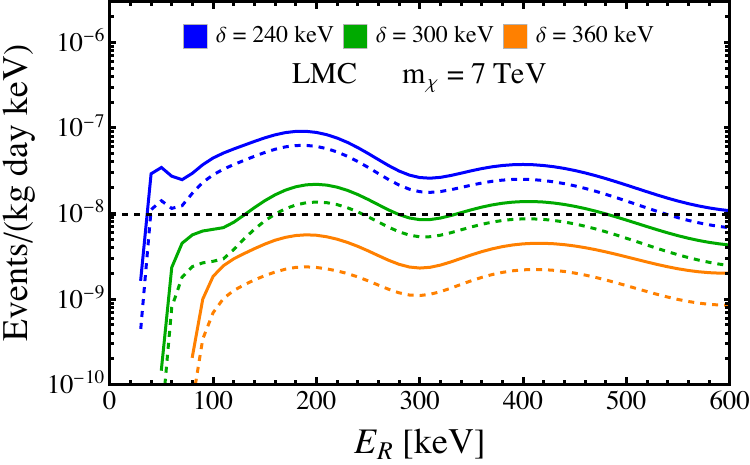}
    }\\[0.75em]
    \resizebox{\columnwidth}{!}{\includegraphics[width=0.5\linewidth]{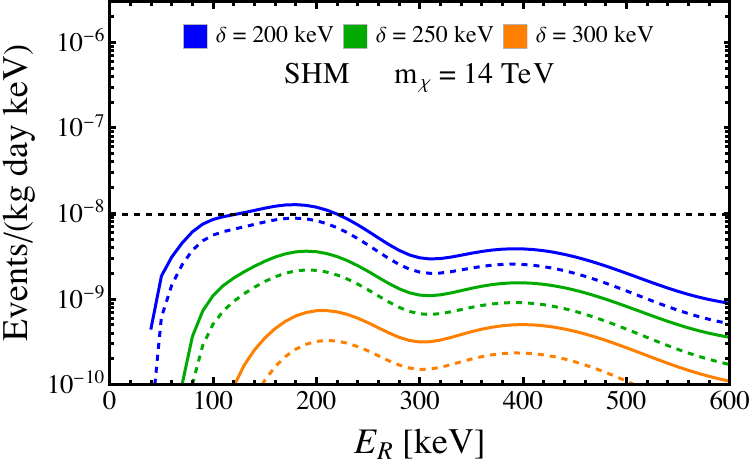}
    \quad
    \includegraphics[width=0.5\linewidth]{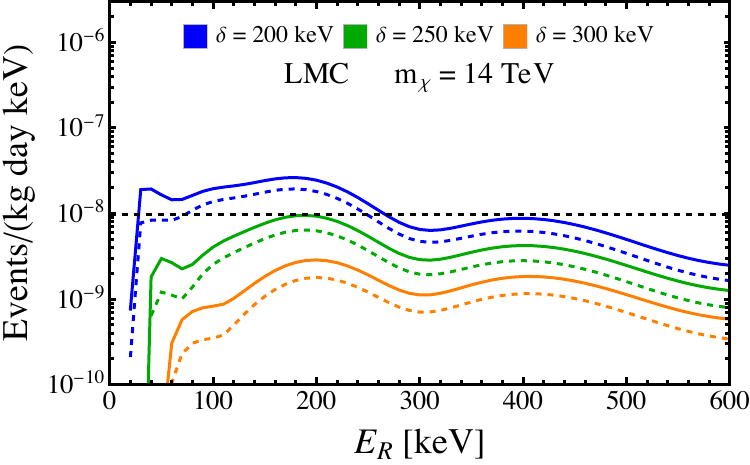}
    }
    \caption{
    Differential recoil spectra for the SHM (\textbf{left}) and MW+LMC (\textbf{right}) velocity distributions introduced in \cref{fig:SHM}, evaluated at illustrative values \(m_\chi=7~\mathrm{TeV}\) (\textbf{top}) and \(14~\mathrm{TeV}\) (\textbf{bottom}). The colors denote the indicated benchmark mass splittings \(\delta\), while the solid (dashed) curves correspond to the maximum (minimum) annual Earth velocity and hence bracket the annual variation. The horizontal black dashed line represents one expected event in a \(100~\mathrm{keV}\) bin at the current LZ exposure. Annual modulation produces only a small variation, whereas the high-velocity tail of the MW+LMC distribution appreciably enhances the rate and extends it to larger mass splittings. Increasing \(m_\chi\) or \(\delta\) generally suppresses the signal.
    }
    \label{fig:spectra_varied_delta}
\end{figure}

\begin{figure}[t]
    \centering
    \resizebox{\columnwidth}{!}{\includegraphics[width=0.5\linewidth]{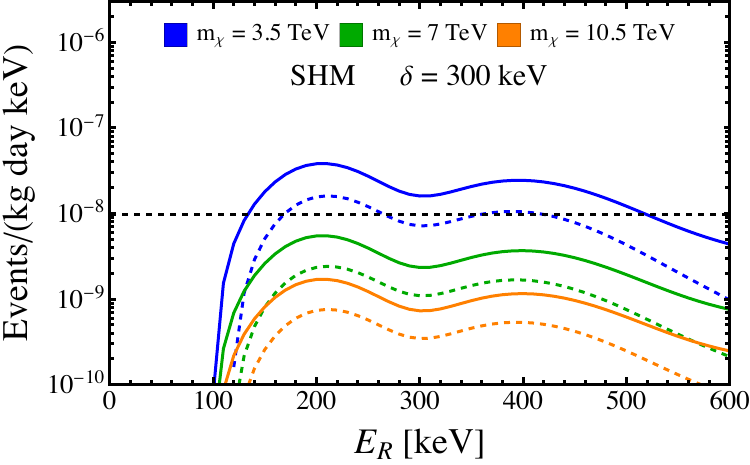}
    \quad
    \includegraphics[width=0.5\linewidth]{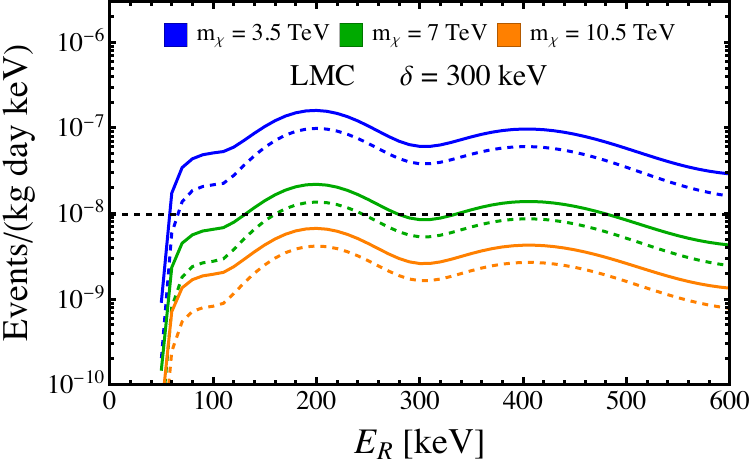}
    }\\[0.75em]
    \resizebox{\columnwidth}{!}{\includegraphics[width=0.5\linewidth]{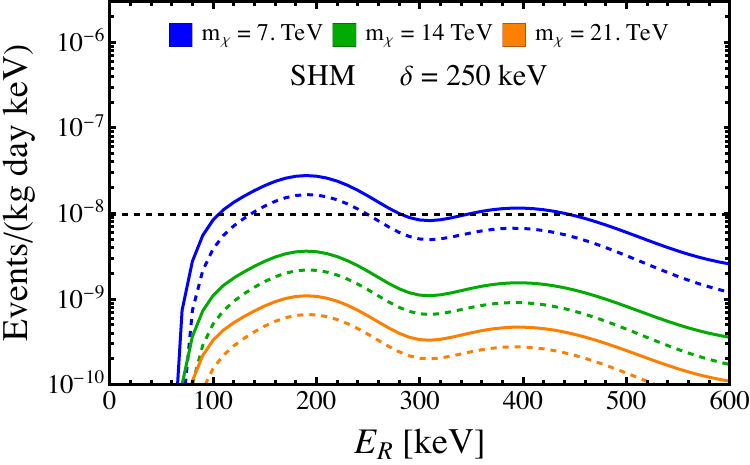}
    \quad
    \includegraphics[width=0.5\linewidth]{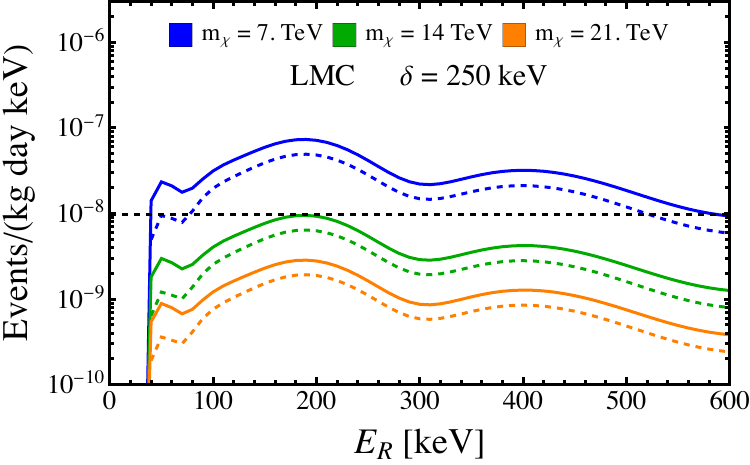}
    }
    \caption{ 
    Differential recoil spectra for the SHM (\textbf{left}) and MW+LMC (\textbf{right}) velocity distributions introduced in \cref{fig:SHM}, evaluated at illustrative values \(\delta=300~\mathrm{keV}\) (\textbf{top}) and \(250~\mathrm{keV}\) (\textbf{bottom}). The colors denote the indicated benchmark dark-matter masses \(m_\chi\), while the solid (dashed) curves correspond to the maximum (minimum) annual Earth velocity and hence bracket the annual variation. The horizontal black dashed line represents one expected event in a \(100~\mathrm{keV}\) bin at the current LZ exposure. Annual modulation has only a small effect compared with the choice of velocity distribution, while the high-velocity tail of the MW+LMC profile appreciably enhances the rate. Increasing either \(m_\chi\) or \(\delta\) generally suppresses the signal. 
    }
    \label{fig:spectra_varied_mchi}
\end{figure}

The differential event rate as a function of the recoil energy $E_R$ is illustrated in \cref{fig:spectra_varied_delta,,fig:spectra_varied_mchi} for different dark matter masses and mass splittings. 
The solid (dashed) lines show the rate obtained with $b = 1/2$ ($-1/2$), capturing the two extremes of the annual modulation.
The panels on the left show the distributions obtained with the SHM while those on the right show the distributions obtained with our modified halo function accounting for the effects of the LMC. 
The dashed line at $\dd R / \dd E_R \approx 10^{-8} / (\textrm{kg}\cdot\textrm{day}\cdot\textrm{keV})$ corresponds to a single event in a 100 keV bin with the exposure of the LZ search~\cite{LZ:2026axp}.

There are several features apparent from the distributions in \cref{fig:spectra_varied_delta,,fig:spectra_varied_mchi}. 
As expected for inelastic scattering, the mass splitting leads to a kinematic lower-cutoff on the recoil energy. 
For larger dark matter masses, once the combined kinematics from the halo and the scattering cross section become unimportant, the total rate scales like $ 1/ m_{\chi}^3$. This is as expected: as the dark matter number density falls like $1/ m_{\chi}$ while for fixed $g_M$, the dipole interaction is inversely proportional to $m_{\chi}$, and this interaction appears squared in the cross section. 

The choice of velocity distribution---SHM or MW+LMC---significantly affects both the total event rate and the low-energy recoil cutoff. 
As shown in the right panel, sufficiently large mass splittings eliminate the rate at low recoil energies for the SHM, while the high-velocity tail of the MW+LMC distribution can sustain an appreciable signal.

The distribution is overall quite flat away from kinematic thresholds: the rate falls less than an order of magnitude between $E_R = 200~\textrm{keV}$ and $500~\textrm{keV}$ for a broad range of masses and mass-splittings. 
This is especially true when using the modified, LMC halo function, which tends to flatten the distribution out to higher recoil energies. The high-velocity tail also extends the {\em maximum} recoil energy to larger values. 
While the general features are not stark, for a broad range of masses above $\sim 500~\textrm{GeV}$ and for mass splittings above $\sim 200~\textrm{keV}$, the rate is largest at a recoil energy $\simeq 200~\textrm{keV}$. 

As a final note, we see again that for most parameters, the two extremes of the annual modulation do not make a large impact on the rate. This difference is even less pronounced compared to the effects of the velocity distribution.
For simplicity, we will neglect this modulation and use the average prediction from each velocity profile in our parameter scan below. We remark only in passing the date the event at LZ was recorded, June 16, is near the peak of the annual modulation. 

\medskip

We can now turn to the total integrated rate in our model.
While only very little information on the event rate distribution can be gleaned from the observation of a single recoil event, we can still understand broadly the range of $m_{\chi}$ and $\delta$ values consistent with the LZ observation~\cite{LZ:2026axp}.
To do so, we integrate the event rate in six, $100~\textrm{keV}$ bins of recoil energy, assuming 220 live days of data and 4.7 tonne of liquid Xenon, as in the LZ search~\cite{LZ:2026axp}.\footnote{
Strictly speaking, the LZ search was only up to $270~\textrm{keV}$. 
There is a gap between $\sim 270$ and $350~\textrm{keV}$, and a separate high-energy sideband from $\sim 350$ to $590~\textrm{keV}$\,NR. This high-energy sideband is used to calibrate the multiple scatter, single ionization background, and not for a dark matter search, but the LZ result reports zero events in this region with their science veto. The importance of this sideband region for inelastic signals, particularly the Higgsino interpretation of the LZ event, was emphasized in particular in Ref.~\cite{Rodd:2026tyn,Dent:2026bji}. 
The efficiency in this region is not public, but is not expected to degrade considerably compared to the science region (per private communication to Rodd et. al),
so we keep these bins in our simple analysis. 
We thank Nick Rodd for private communication on this topic.
} 
We then compute the extended negative log-likelihood ($\mathcal{L}$) of the single event observation in the $200 - 300~\textrm{keV}$ bin, neglecting the small Standard Model background for simplicity. 
This accounts for the important features that our MIDM model must predict: the absence of events at low recoil, which constrains the mass splitting, and the overall event rate, which depends strongly on both $m_{\chi}$ and $\delta$, the latter due to the effect on the minimum velocity as a function of $E_R$, as well as the large uncertainty on the exact recoil energy of the event.

\begin{figure}[p]
    \centering
    \includegraphics[width=0.85\linewidth]{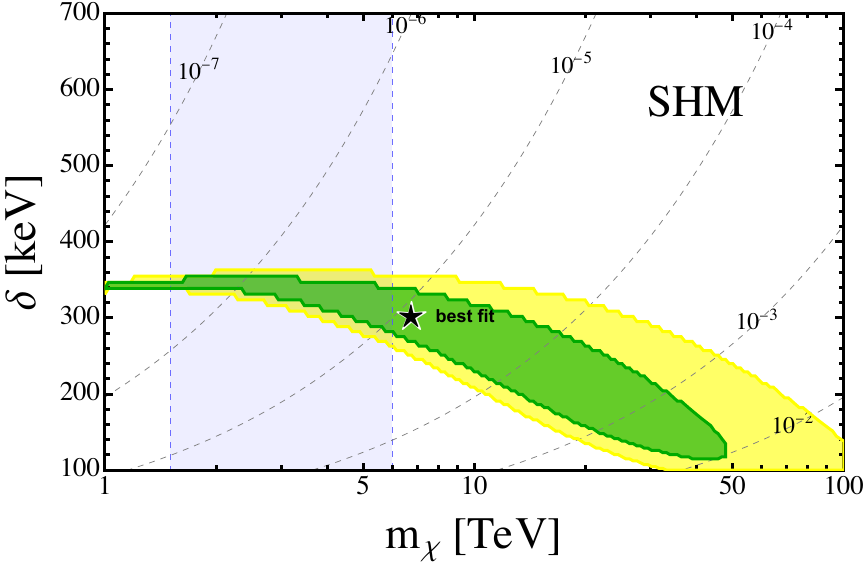} \\[0.5em]
    \includegraphics[width=0.85\linewidth]{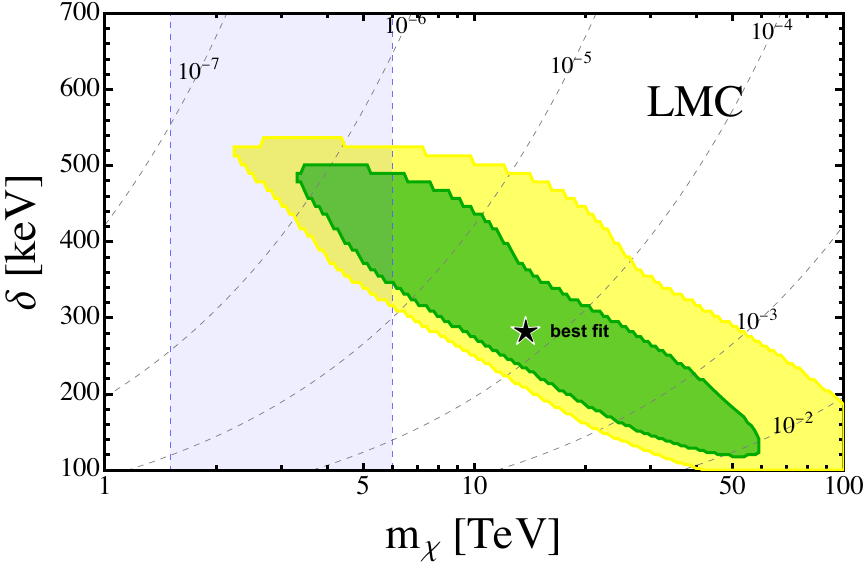}
    \caption{
    The \(1\sigma\) (\textcolor{GoodGreen}{green}) and \(2\sigma\) (\textcolor{GoodYellow}{yellow}) regions around the best-fit point to the LZ high-recoil event, obtained using the SHM (\textbf{top}) and MW+LMC (\textbf{bottom}) velocity distributions from \cref{fig:SHM}. The \textcolor{LighterBlue}{light-blue} band denotes the range of \(m_\chi\) for which the heavy-quark regime of our model predicts a sufficiently small mass splitting \(\delta\) (see \cref{fig:massSplitting}). Dashed \textcolor{gray}{gray} lines show contours of constant excited-state lifetime in seconds.  For typical velocities of the excited state, we anticipate a significant fraction, $\mathcal{O}(1)$ for $\tau < 10^{-6}$ and $\mathcal{O}((10^{-6} \; {\rm s)/\tau)}$ for $\tau > 10^{-6}$, of nuclear recoil events will also contain the excited state decay photon depositing electron-equivalent energy inside the detector; see the text for details.
    }
    \label{fig:param_scan}
\end{figure}

We then draw contours in the mass splitting vs.~dark matter mass plane that correspond to $1\sigma(2\sigma)$ preferred regions, obtained by computing $-2\Delta \mathcal{L} = 2.3 (6.2)$ around the minimum value. 
This is shown in \cref{fig:param_scan} for two different dark matter velocity profiles.
The best-fit points are \((m_\chi,\delta)=(6.8~\mathrm{TeV},300~\mathrm{keV})\) for the SHM and \((14~\mathrm{TeV},280~\mathrm{keV})\) for the modified MW+LMC profile, close to the benchmarks used in \cref{fig:spectra_varied_delta,fig:spectra_varied_mchi}.  As can be seen in these figures, at these parameter points the rate is broadly distributed across recoil energy, with the predicted number of events in each $100~\textrm{keV}$ bins from zero to 600~keV  
$\{ 0, 0.2, 0.3, 0.2, 0.2, 0.1 \}$ and $\{ 0, 0.3, 0.3, 0.2, 0.2, 0.1\}$, respectively.

Much of the preferred region can be understood by the kinematics imparted by the mass splitting, which must be large enough to forbid too many events at low recoil, but not too large or it suppresses the total scattering rate due to the falling halo function at high velocities. 
Assuming the Standard Halo Model, this results in a very narrow range of $\delta$ values that are compatible with the signal.
At higher masses, the rate falls steeply, as anticipated above, and this must be compensated by smaller values of the mass splitting. This results in the inverse correlation between $\delta$ and $m_{\chi}$ visible in both panels. At large masses, the preferred region in $\delta$ broadens significantly, as the effects of the kinematics are less important on the total rate. 
We see that, depending on the mass splitting and assuming the SHM, the LZ event is consistent with a broad range of dark matter masses, ranging from $\sim 1$ to $50~\textrm{TeV}$. 
This is easily compatible with the motivated region illustrated by a blue band for which we obtain mass splittings of $\mathcal{O}(100~\textrm{keV})$ in our particular realization of strongly-interacting magnetic inelastic DM, assuming the heavy quark limit as discussed in \cref{sec:mass_splitting}.

Taking the modified velocity profile that accounts for the effects of the LMC, however, we find a significantly different preferred region.
As seen above, the high-velocity tail tends to flatten the recoil energy distribution, so there is no strong preference over a broader range of masses and mass splittings.
The high-velocity tail also allows for a much larger mass splitting to be compatible with the observed event. 
On the other hand, both larger and smaller energy recoils are more expected, and this makes the fit worse for lighter dark matter masses, so the $1 \sigma$ preferred region extends only down to $\sim 3~\textrm{TeV}$. 

We emphasize that these preferred regions should not be over interpreted, as they are generated with a crude statistical test of the distribution, neglect the backgrounds, and all non-statistical uncertainties. 
In fact, the difference between the values preferred by the SHM and MW+LMC velocity profiles illustrates the significant uncertainty in interpreting the observation in the context of any model due to the uncertainty in the velocity profile alone.  
These two profiles give an indication of how uncertainty in the dark-matter velocity distribution impacts the preferred parameter space of our model, motivating further study of the Galactic halo velocity distribution. Encouragingly, in both cases the $(m,\delta)$ region (\textcolor{LighterBlue}{light blue}) discussed in Section~\ref{sec:noblemodels} overlaps substantially with the region favored by the experimental data. 
This rudimentary analysis is enough to demonstrate that a confining dark sector could explain the observed event for a broad range of masses in the $\sim \textrm{few}$ to $10$s of TeV ballpark. 

In all the results presented above, we have assumed $g_M = -2 \sqrt{3}$ as in \cref{eq:g_M_numerical}. 
Aside from this value, the results in \crefrange{fig:spectra_varied_delta}{fig:param_scan} apply equally well to any magnetic inelastic dark matter model, assuming the other interactions are sufficiently suppressed. 
For most of the parameter space, the rate scales as $g_M^2 / m_{\chi}^3$, so $\mathcal{O}(1)$ different values of $g_M$ can be extrapolated from our results. 
Significantly smaller values of $g_M$ that would result from e.g., a weakly coupled UV completion, would look significantly different due to the kinematic dependence of the recoil energy distributions.

%%%%%%%%%%%%%%%%%%%%%%%%%%%%%%%%%%%%%%%%%%%%%%%%%%%%%%%%%%%%%%%%%%%%%
\section{Discussion}

We have shown that dark-baryon dark matter with a transition magnetic dipole moment can account for the LZ high-recoil event through inelastic scattering. In the confining models of Refs.~\cite{Asadi:2024bbq,Asadi:2024tpu}, dark quarks charged under \(\mathrm{SU}(2)_L\) form neutral baryons with opposite \(\Hp\)-parities. This symmetry forbids their diagonal one-photon electromagnetic moments while allowing a transition moment, suppressing the more strongly constrained elastic signal. For three dark colors and three flavors transforming as an electroweak triplet, the heavy-quark limit can also produce a sufficiently small splitting between the neutral triplet and quintuplet baryons.

A detailed calculation of the recoil spectrum shows that multi-TeV dark matter with \(\delta\sim\mathcal{O}(100s)\,\mathrm{keV}\) can reproduce the observed event. Although the preferred masses and splittings depend appreciably on the assumed halo velocity distribution, viable regions are found using both the SHM and MW+LMC profiles. Annual modulation has a comparatively small effect relative to this halo uncertainty.

This framework also predicts correlated signatures beyond the observed recoil. Below, we comment on implications of this setup at different experiments.

\subsection*{Direct Detection}

Strongly-coupled magnetic inelastic dark matter predicts an array of correlated direct detection signals.
In this discussion, we will consider both $B_3^0$ and $B_5^0$ as potential dark matter candidates, since we have seen from \cref{sec:mass_splitting} that either of these states could be the lightest dark baryon.
\begin{itemize}
\item \textbf{Elastic scattering}~~
As emphasized in Ref.~\cite{Asadi:2024bbq}, we expect elastic scattering of the dark matter with nuclei.  There are three sources of this:  1) Because $B_3^0$ and $B_5^0$ are embedded in (3-plet and 5-plet) electroweak representations, they will have electroweak loop-induced contributions to elastic scattering, calculated using an EFT approach in Ref.~\cite{Chen:2023bwg}.  The absence of a direct detection signal at lower recoil energies suggests $m_{B_{3,5}^0} \gtrsim \mathcal{O}(100s \; {\rm GeV})$. 
2) Polarizability, the first allowed electromagnetic moment by $\mathcal{H}$-parity, is expected to be present for both of these candidates as well as a candidate arising from the global flavor model. Estimates for the size of the dark baryon polarizability and the relevant nuclear matrix elements include Refs.~\cite{Weiner:2012cb,Ovanesyan:2014fha,Appelquist:2015zfa,Brod:2017bsw,Kavanagh:2018xeh}. These estimates were used in Ref.~\cite{Asadi:2024bbq} where it was found to be even more suppressed compared with the electroweak loop-induced contribution.
3) Standard Model interactions violate $\mathcal{H}$-parity, leading to loop-induced contributions to the dark baryon magnetic dipole moments.  Again, the estimates from \cite{Asadi:2024bbq} suggest these are also even more suppressed than the electroweak loop-induced contributions.
\item \textbf{Dark matter upscatter and decay}~~ The excited state dark baryon, $\chi_h$, will decay back to the dark matter ground state $\chi$ and a photon with width \cite{Eby:2023wem}
\begin{eqnarray}
\label{eq:excitedwidth}
  \Gamma_{\chi_h} \;\equiv\; \frac{1}{\tau_2} &=& \frac{ \mu_\chi^2}{\pi} \delta^3 \, , 
\end{eqnarray}
leading to a typical decay length given by
$\ell_{\chi_h}$ 
\begin{equation}
\label{eq:decaylength}
  \ell_{\chi_h} \;=\; v \, \tau_2 \; \simeq \;
 \left( 4.7 \; {\rm m} \right) \times 
 \left(\frac{2 \sqrt{3}}{g_M}\right)^2\left(\frac{m_{\chi_h}}{5 \; {\rm TeV}}\right)^2
 \left(\frac{300\,{\rm keV}}{\delta}\right)^3
 \left(\frac{v}{700\,{\rm km/sec}}\right) \, .
\end{equation}
This distance is comparable to the size of dark matter detector itself, allowing for the possibility of seeing \emph{both} the nuclear recoil and the excited state decay in the same event.  This idea was originally studied in Refs.~\cite{Chang:2010en,Lin:2010sb}, where they emphasized that a nuclear recoil followed by a decay predicts two different signals simultaneously in the same event -- one consistent with a nuclear recoil with large $E_R$, and the second signal is a photon with energy $\delta \sim 100$-$500$~keV consistent with Fig.~\ref{fig:param_scan}.  Moreover, \cite{Lin:2010sb} emphasized that the nuclear recoil plus decay signal has a directional dependence.  This directional dependence relies on the dark matter being (much) heavier than the nucleus, so that the excited state is preferentially heading in the same direction after scattering off the nuclei.
Hence, this ``double scatter'' (nuclear recoil and photon deposition) signal can not only be used to separate the dark matter detection signal from the background,\footnote{Direct detection experiments have routinely rejected events that have signals consistent with multiple scattering (multiple independent S1/S2 events).  For example, the recent LZ analysis  specifically requires ``events are selected
as signal candidates if they are classified as single scatters'' \cite{LZ:2026axp}, and thus would require a new search strategy to find these double scatter events.} but it also allows us to learn about the dark matter velocity distribution itself. 
\item \textbf{Dark matter decay-to-photon only signals}~~ Magnetic inelastic dark matter can also be observed just using the electromagnetic signal from the excited state decay. The idea was originally proposed in Ref.~\cite{Feldstein:2010su} in the context of explaining the DAMA annual modulation signal, and later developed in Refs.~\cite{Pospelov:2013nea,Eby:2019mgs,Eby:2023wem,Graham:2024syw}.  One of the striking possibilities is a sidereal-daily modulating rate of the photon signal resulting from scattering off the rock in the Earth \cite{Feldstein:2010su,Eby:2019mgs,Eby:2023wem} or additional material added surrounding the detector \cite{Pospelov:2013nea,Graham:2024syw}.

\end{itemize}

\subsection*{Collider Signals}

\begin{itemize}
\item \textbf{Pseudoscalar mesons}~~ In the light-quark limit, where the chiral Lagrangian is valid, dark pseudoscalar mesons can lead to disappearing tracks via cascade decays or electroweak diboson resonances via chiral anomalies \cite{Asadi:2025vfr}. Existing searches \cite{ATLAS:2022rme,CMS:2023mny,ATLAS:2018iui,ATLAS:2020tlo,ATLAS:2020fry,CMS:2021wlt,CMS:2021klu,ATLAS:2021uiz,ATLAS:2023sua,CMS:2024nht,CMS:2026xbb} for these signatures already probe $\mathcal{O}(1)\,$TeV-scale dark mesons, so some parameter space for inelastic dark baryon DM consistent with the LZ event may be already probed or soon within reach. 

\item \textbf{Vector mesons}~~ The dark $\rho$ meson may also act as a $Z'$-like resonance in colliders that decays to leptons via mixing with the Standard Model $Z$. Dilepton resonance searches \cite{CMS-PAS-EXO-25-021,ATLAS:2026nqc} probe the several-TeV scale, but a mixing angle of $\sim\!0.01$ could evade bounds on a Sequential Standard Model $Z'$. We leave estimation of this mixing angle to future work, as there are subtleties due to dependence on $N_c$ and $N_f$ (see, e.g., \cite{Kribs:2018ilo}), as well as non-perturbative uncertainties that become difficult to control away from the light-quark limit.

\item \textbf{Quirks}~~ In the heavy-quark limit, the dark sector may exhibit quirk-like collider signals \cite{Knapen:2016hky}. When $\Lambda_\chi\ll m_q$, a dark $q\bar{q}$ pair that is joined by an oscillating dark color flux tube may be produced rather than well-defined dark hadrons. The space of possible phenomenology is vast (see Refs.~\cite{Forsyth:2025wks,Curtin:2025ngf} for recent examples and Ref.~\cite{Asadi:2026mip} for additional discussion), but existing searches at Tevatron \cite{D0:2010kkd} and FASER \cite{FASER:2026fsy} are most sensitive to the ``macroscopic'' quirk limit with $\Lambda_\chi\lesssim1\,$MeV and $m_q$ of $\mathcal{O}(100)\,$GeV. A discovery of quirks would not only be suggestive of dark baryon DM but also have implications for the reheating temperature of the early universe due to dark glueball decays \cite{Asadi:2025btr}.

\end{itemize}

\subsection*{Astrophysical Searches}

\begin{itemize}
\item \textbf{Solar capture}~~ A recent analysis \cite{Pospelov:2026ewn} has demonstrated there are strong constraints from IceCube \cite{IceCube:2025fcu} on the inelastic Higgsino interpretation of the LZ observation, arising from the accumulation of dark matter in the Sun and the subsequent annihilation into $W$-pairs that decay into neutrinos.   If the mechanism to generate the relic abundance of the dark baryon dark matter considered in this paper results in a symmetric abundance of dark baryons and anti-baryons, we expect that dark baryon capture and annihilation in the Sun can also occur in our models.  However, there are several differences between an inelastic Higgsino model and the strongly-coupled magnetic inelastic model.  Firstly, the dominant scattering process for the dark baryons is through a spin-dependent nuclear response, and there is a much smaller fraction of heavier elements in the Sun with nonzero spin.  For example, the dominant capture elements identified in Ref.~\cite{Pospelov:2026ewn} include iron, that contains only 2.2\% of ${}^{57}$Fe, and nickel, that contains only 1.1\% of ${}^{61}$Ni.  
Secondly, the annihilation of dark baryons with dark anti-baryons is anticipated to have many additional strong channels, where the annihilation proceeds into meson--anti-meson pairs.  Using QCD as an example, the annihilation rate of proton--anti-proton into photons is less than $10^{-6}$ relative to the annihilation rate into mesons.  In practice, we need to take into account kinematic suppression (e.g., in the heavy quark limit, $m_{\rm meson} \simeq 2\,m_{\rm baryon}/N_c $), and the subsequent decay of the mesons themselves.  If the dominant annihilation channel of dark baryons were to the triplet of pions, the charged dark pions decay to the neutral dark pion emitting a very soft $\pi^\pm$ \cite{Asadi:2025vfr}.  The neutral pion is protected by $G$-parity \cite{Bai:2010qg}, so it is either stable (an additional component of dark matter), or unstable due to $G$-parity violating interactions that need not have any decays that result in final state neutrinos.  This suggests the fraction of dark baryon--anti-baryon annihilation into energetic neutrinos has substantial suppression compared with an elementary WIMP candidate.
We conclude that solar capture and annihilation into neutrinos is not obviously a significant constraint on the models presented in this paper, due to the smaller capture rates and the model-dependence of the annihilation rates.

\item \textbf{Galactic center and dwarf galaxies}~~ A qualitatively different result may arise from annihilation in the galactic center or dwarf galaxies.  Unlike the capture and annihilation in the Sun, the total rate for annihilation in the galactic center or dwarfs is the sum of the rates of dark baryon annihilation into Standard Model final states as well as the larger rates into mesons discussed above.  Assuming a symmetric abundance of dark baryons and anti-baryons, the annihilation rates into Standard Model final states are thus likely to be comparable to elementary Majorana WIMP candidates, with the same SU(2)$_L$ representation, that are the traditional benchmarks for indirect detection.
As a member of an electroweak multiplet, the dark matter can annihilate into Standard Model gauge bosons through the exchange of charged dark baryons in the \(t\)-channel, producing indirect-detection signals analogous to those of minimal electroweak multiplet dark matter.

\item \textbf{Sommerfeld enhancements}~~ Sommerfeld enhancement and bound state effects are studied for these relics in details \cite{Hisano:2004ds,Arkani-Hamed:2008hhe,Cirelli:2015bda,Garcia-Cely:2015dda,Asadi:2016ybp,Mitridate:2017izz,Baumgart:2017nsr,Rinchiuso:2018ajn,Baumgart:2018yed,Baumgart:2023pwn,Baumgart:2025dov} and can drastically modify their indirect detection signals. 
Sommerfeld enhancement, in particular, can substantially increase the indirect-detection signal from heavy electroweak relics, making the sensitivity strongly dependent on whether the dark-matter mass lies near a resonance or in a minimum between resonances. The strongest current constraints in the multi-TeV range come from the H.E.S.S. search for gamma-ray lines from the Galactic Center that used an inner galactic dark matter density profile that is strongly centrally concentrated. For canonical elementary \(Y=0\) Majorana multiplets with radiatively generated charged–neutral mass splittings, the H.E.S.S. limits obtained using an Einasto halo profile probe triplet and quintuplet dark matter up to approximately \(5\) and \(20~\mathrm{TeV}\), respectively \cite{HESS:2026ila}---see the right panel of Fig.~1 therein.
Isolated Sommerfeld resonances may extend the sensitivity to substantially larger masses, whereas the intervening minima leave weaker or unconstrained regions.  

These limits, however, cannot be applied directly to the composite dark baryons considered here. First, the Sommerfeld potential is controlled by the splittings between the neutral and charged members of each electroweak multiplet, rather than by the small \(B_3^0\)–\(B_5^0\) splitting relevant for inelastic direct detection. The resonance locations must therefore be recalculated using the charged-baryon spectrum of our model. Moreover, the H.E.S.S. predictions assume elementary Majorana fermions, while the spin, statistics, and particle–antiparticle structure of the composite baryons modify both the available two-body channels and the normalization of the annihilation rate. Their short-distance annihilation amplitudes may also receive non-perturbative contributions from the confining dynamics, as well as composite form-factor corrections, and consequently need not coincide with the electroweak annihilation matrices used for elementary triplets and quintuplets. Finally, the Galactic-Center bounds depend strongly on the assumed dark-matter density profile: relative to the baseline Einasto result, the H.E.S.S. limit weakens by approximately a factor of \(5.8\) for the cored-NFW profile considered in the same analysis. 
The quoted reaches should therefore be regarded as indicative benchmarks for indirect detection, rather than as exclusions of the corresponding mass ranges in our model.
The mass range favored by the LZ result is accessible to astrophysical searches for the DM annihilation, at telescopes such as H.E.S.S\@.  Improvements in the determination of astrophysical inputs, such as $J$-factors, will be critical to determining the indirect detection signal.

\item \textbf{Dark nuclei}~~ 
It is possible that dark nuclei form in the early universe and are themselves stable, providing an additional subcomponent of dark matter.  In a confining SU(2) theory \cite{Detmold:2014qqa}, dark nuclei were identified and studied.  Novel indirect detection signals could result from dark nucleosynthesis occurring within stars. 
In \cite{Redi:2018muu}, the cosmological production of dark nuclei was computed specifically for dark baryons that transform as triplets of SU(2)$_L$.  Their conclusion was that only a fraction of dark matter binds into dark deuterium, and a negligible amount of dark tritium.  Examining in more detail what fraction would be expected within the parameter space identified in this paper is worthy of a future study.

\end{itemize}

\subsection*{Open Questions in the Dark Sector}
\begin{itemize}
    \item \textbf{Relic abundance}~~ The astrophysical constraints discussed above assume a symmetric dark-matter abundance and can be substantially weakened if the relic is asymmetric. The dark-baryon abundance may be determined by several processes, including annihilation into Standard Model particles or lighter unstable dark mesons. Moreover, in the heavy-quark regime considered here---see \cref{fig:massSplitting}—thermal squeeze-out \cite{Asadi:2021yml,Asadi:2021pwo,Asadi:2022vkc} will efficiently deplete the symmetric dark-baryon component. 
    An additional mechanism to generate a primordial asymmetric abundance will then be required to reproduce the observed abundance. 
    The resulting scarcity of dark antibaryons would suppress annihilation signals both from the Galactic halo, weakening the H.E.S.S. limits, and from dark matter captured in the Sun. We leave a detailed calculation of the relic abundance and its implications for these constraints to future work.

    \item \textbf{Other representations}~~ Finally, it is worth pointing out the model in this paper with three dark colors and three dark quark flavors is the simplest of the Noble Dark Matter models \cite{Asadi:2024tpu} that has dark baryons dark matter candidates in adjacent SU(2)$_L$ representations that can give rise to the transition magnetic moment responsible for the inelastic scattering signal. 
    Similar phenomenology can arise from models with other larger numbers of dark colors and dark quark flavors and we will their phenomenology to future works.
\end{itemize}

%%%%%%%%%%%%%%%%%%%%%%%%%%%%%%%%%%%%%%%%%%%%%%%%%%%%%%%%%%%%%%%%%%%%%
\section*{Acknowledgments}

We are grateful to Eric Dahl, Matthew Reece, Nick Rodd, and Neal Weiner for helpful discussions. 
The work of PA is supported in part by the U.S. Department of
Energy grant number DE-SC0010107.
AB is supported by the Walter Burke Institute for Theoretical Physics. 
PF is supported by Fermilab which is administered by
Fermi Forward Discovery Group, LLC under Contract
No. 89243024CSC000002 with the U.S. Department of
Energy, Office of Science, Office of High Energy Physics.
The work of GK was supported in part by the U.S. Department of Energy under grant number DE-SC0011640.

%%%%%%%%%%%%%%%%%%%%%%%%%%%%%%%%%%%%%%%%%%%%%%%%%%%%%%%%%%%%%%%%%%%%%
\bibliographystyle{utphys}
\bibliography{lz-refs}

\end{document}